\documentstyle[manuscript,aps,prbbib]{revtex}

\tighten
\newcommand{\be}{\begin{equation}}
\newcommand{\ee}{\end{equation}}
\newcommand{\bea}{\begin{eqnarray}}
\newcommand{\eea}{\end{eqnarray}}
\newcommand{\bd}{\begin{displaymath}}
\newcommand{\ed}{\end{displaymath}}
\newcommand{\bi}{\begin{itemize}}
\newcommand{\ei}{\end{itemize}}
\newcommand{\bc}{\begin{center}}
\newcommand{\ec}{\end{center}}
\newcommand{\bfl}{\begin{flushleft}}
\newcommand{\efl}{\end{flushleft}}
\newcommand{\bfr}{\begin{flushright}}
\newcommand{\efr}{\end{flushright}}
\newcommand{\f}{\frac}

\def\6{\partial} \def\a{\alpha} 
 \def\d{\delta}

\def\o{\omega}  
 \def\L{\Lambda}

\def\={\!\!\!&=&\!\!\!}
\def\+{\!\!\!&&\!\!\!+~}
\def\-{\!\!\!&&\!\!\!-~}

\begin{document}

\title{Renormalization group approach of itinerant electron systems near the
Lifshitz point}
\author{C.P. Moca, I. Tifrea and M. Crisan}
\address{Department of Theoretical Physics\\
University of Cluj, 3400 Cluj, Romania}
\maketitle

\begin{abstract}
Using the renormalization group approach proposed by Millis for the
itinerant electron systems we calculated the specific heat coefficient
$\gamma(T)$ for the magnetic fluctuations with susceptibility
$\chi^{-1}\sim |\d+\o|^\a+f(q)$ near the Lifshitz point. The constant value
obtained for $\a=4/5$ and the logarithmic temperature dependence, specific for
the non-Fermi behavior, have been obtained in agreement with the experimental
data.
\end{abstract}

The occurrence of the non-Fermi behavior in the systems of fermions coupled
to a critical fluctuations mode has been suggested in connection with the
neutron experiments \cite{1,2} and studied in the framework of many body
theory \cite{3,4} in the case of two-dimensional (2D) and three dimensional
(3D) models.

Recent experiments on the heavy fermions systems showed also a non-Fermi
behavior of these materials at low temperatures and it was associated with
the proximity of quantum critical point (QCP). The most studied example
\cite{5} is CeCu$_{6-x}$Au$_x$ where at the QCP $x=0.1$ the resistivity
increases linearly with temperature T over a wide range of T and the specific
heat C(T) is proportional to $T\ln{T_0/T}$. This behavior has been explained
\cite{6} by the coupling of 3D fermionic excitations to the 2D critical
ferromagnetic fluctuations near the QCP.

The inelastic neutron scattering measurements performed on this materials
\cite{7,8} showed the following new points in the behavior of this material
\begin{enumerate}
\item The inelastic neutron scattering data can be fitted with a susceptibility
of the form $\chi^{-1}=C^{-1} [f(q)+(aT-i\o)^\a]$ where $\a=4/5$ and not 1 as is
predicted by the mean field approximation
\item The quadratic stiffness vanishes, fact which shows that we are dealing
with a quantum Lifshitz point (QLP)
\item The peaks for $x=0.2$ and $x=0.3$ can be considered as 2D precursor of
3D order
\item The scaling analysis showed \cite{7} that $\gamma(T)=C(T)/T$ has the
form
\be
\gamma(T)\sim T^{(D-1/2)\a/2-1}
\label{e1}
\ee
which for $D=3$ and $\a=4/5$ gives a temperature independent value.
\end{enumerate}
This analysis has been performed taking as the most important
contribution to $\chi$ the form containing $\o^\a$ and the q-dependence of the
form $f(q)=D q^2_\parallel+C q^4_\perp$ where $\a=2/z$, z being the critical
exponent from the dynamical critical phenomena \cite{9}.

In this paper we will show that using the Hertz \cite{10} renormalization
group method (RNG) extended for $T\neq 0$ by Millis \cite{11} we can obtain
the $\ln{T_0/T}$ term as a quantum correction to the classical results
expressed by Eq. (\ref{e1}). We start from an interacting Fermi system and by
introducing the Bose field $\Phi(q)$ (associated with the magnetic fluctuations)
via the Hubbard-Stratanovich transformations one can integrate out the fermions
and expand the effective action up to forth order in $\Phi$ as
\bea
S_{eff}[\Phi]&=&\int_0^1\f{d\o}{2\pi}\int_0^1\f{d^2q_\parallel}{(2\pi)^2}
\int_0^1\f{dq_\perp}{2\pi}\Phi\left[\left|\d+|\o|\right|^\a+q_\parallel^2+
D q_\perp^2+q_\perp^4\right]\Phi\nonumber\\
&+&u\int_0^1\f{d\o}{2\pi}\int_0^1\f{d^2q_\parallel}{(2\pi)^2}\int_0^1
\f{dq_\perp}{2\pi}(\Phi \Phi)^2
\label{e2}
\eea
where $u>0$ is the coupling constant. The scaling variables of the models are
$\d^\a$ ($\d$ is the deviation of the control parameter $x$ from its critical
value $x_c$) the stiffness D, the temperature and the coupling constant $u$.
Using the transforms
\bea
\o'=b^{2/\a}\o \hspace{1cm} q'_\parallel=b q_\parallel \hspace{1cm}
q'_\perp=q_\perp \sqrt{b}\nonumber\\
\d'=b^{2/\a}\d \hspace{1cm} D'=bD
\label{e3}
\eea
and interacting out on the shell $\L\geq q\geq\L/b$ ($b>1$) near the cut-off
$\L$ we obtain the RNG equations. In order to calculate the specific heat we
will use the scaling procedure to obtain an equation for the free energy,
defined for the free fluctuations as
\be
F=\int_0^1\f{dz}{2\pi}\int_0^1\f{d^2q_\parallel}{(2\pi)^2}\int_0^1
\f{dq_\perp}{2\pi}\coth{\f{z}{2T}}\arctan{\f{A\sin{\theta}}{A\cos{\theta}+
q^2_\parallel+q_\perp^4}}
\label{e4}
\ee
where $\theta=\a\tan^{-1}{(z/\d)}$ and $A^{-\a}=(\d^2+z^2)^{1/2}$. Following
the same procedure as in Ref. \onlinecite{11} we obtain the equations
\be
\f{dT(b)}{d\ln{b}}=\f{2}{\a} T(b)
\label{e5}
\ee
\be
\f{du(b)}{d\ln{b}}=\left(\f{3}{2}-\f{2}{\a}\right)u(b)-u^2(n+\d)f_2
\label{e6}
\ee
\be
\f{d\d^\a(b)}{d\ln{b}}=2\d^\a(b)+2u(b)(n+2)f_1
\label{e7}
\ee
\be
\f{dD(b)}{d\ln{b}}=D(b)
\label{e8}
\ee
\be
\f{dF(b)}{d\ln{b}}=\left(\f{2}{\a}+\f{5}{2}\right)F(b)+f_3
\label{e9}
\ee
where $n$ is the number of the field components, $f_1=f_1[T(b),\d^\a(b),D(b)]$,
$f_2=f_2[T(b),\d^\a(b),D(b)]$ and $f_3=f_3[T(b)]$ are complicated functions
but will be approximated as presenting a weak dependence of $\d^\a(b)$ and
$D(b)$ for $\d^\a(b), D(b)\ll 1$. The renormalization procedure is stoped at
\be
\d^\a(b)=1
\label{e10}
\ee
and from Eqs. (\ref{e5})-(\ref{e7}) we get
\be
T(b)=T b^{2/\a}
\label{e11}
\ee
\be
u(b)=u b^{3/2-2/\a}
\label{e12}
\ee
\be
\d^\a(b)=e^{2x}\left[\d^\a+2u(n+2)\int_0^{\ln{b}} dx e^{-x(1/2+2/\a)}
f_1\left(Te^{2x/\a}\right)\right]
\label{e13}
\ee
These equations will be analyzed in two regimes. First regime defined by
\be
T(b)\ll 1
\label{e14}
\ee
will be called "quantum regime" and the second defined by
\be
T(b)\gg 1
\label{e15}
\ee
will be called "classical regime". Eqs. (\ref{e5})-(\ref{e8}) have been
solved following Ref. \onlinecite{11} in the quantum and classical regimes
and the renormalization coupling constant has been obtained as
\be
u(b)=uT^{1-3\a/4}\f{T^{3\a/4}}{\left[\bar{\d}^\a+(A-4B)(n+2)uT^{1+\a/4}\right]^3/2}
\label{e16}
\ee
where $\bar{\d}^\a=\d^\a(\bar{b})$ and $\bar{b}$ is defined by $T(\bar{b})=1$
and is $T^{-\a/2}$. The coupling constant have to satisfy the condition
$u(b)\ll 1$ and this condition is not satisfied if
\be
\bar{\d}^\a+(A-4B)(n+2)uT^{1+\a/4}=0
\label{e17}
\ee
If we define the coherence length by $\xi^{-2}\sim \d$ we get
$\xi^{-2}\sim T^{\a+1/4}$.

In order to calculate the specific heat and $\gamma=C(T)/T$ we
will use Eq. (\ref{e9}) for the free energy F. Neglecting in
the lower approximation the second term we obtain
\be
F(T)=F(b) b^{-2/\a-5/2}
\label{e18}
\ee
The exact solution of Eq. (\ref{e9}) has the form 
\be
F(b)=b^{2/\a+5/2}\int_0^{\ln{b}}dx e^{-(2/\a+5/2)x}f_3\left(Te^{2x/\a}\right)
\label{e20}
\ee
and in order to perform the integral in this expression we take the variable
$x$ in the domains
\be
0<x<\f{\a}{2}\ln{\f{1}{T}}
\label{e21}
\ee
\be
\f{\a}{2}\ln{\f{1}{T}}<x<\ln{b^*}
\label{e22}
\ee
where $b^*$ is defined as in Ref. \onlinecite{11}, by $\d(b^*)=1$.
In the first domain \cite{11} $f_3(T)\cong C_3 T^2$ and in
the second $f_3(T)\approx D T$. Using these approximations we obtain from
Eq. (\ref{e20})
\be
b^{-2/\a-5/2} F(b)=\f{\a}{2}T^{1+5\a/4}\left[C_3\int_T^1 dT_1 T_1^{-5\a/4}
+D \int_1^{T b^{*2/\a}} dT_1 T_1^{-1-5\a/4}\right]
\label{e23}
\ee
where $T_1=T\exp{[2x/\a]}$. If we take $\a=4/5$ from Eq. (\ref{e23}) we calculate
\be
F(T)=\f{2}{5}C_3 T^2\ln{\f{1}{T}}+\f{2}{5} D T^2 -\f{2}{5} D T b^{*-5/2}
\label{e24}
\ee
and
\be
\gamma(T)\cong \gamma_{c0}+\bar{\gamma}\ln{\f{1}{T}}+O\left(\f{1}{T^2}\right)
\label{e25}
\ee
a result which shows that the constant value from Ref. \onlinecite{7} is
reobtained but the RNG equations give also the second term specific to
the non-Fermi behavior.

Recently, Ramazashvili \cite{12} used the same method studying QLP for such
a model with $\a=1$. Our results are consistent with the results obtained in
Ref. \onlinecite{12} but the specific heat coefficient obtained has a $T^{1/4}$
dependence which is given by the simple gaussian approximation and the value
$\a=1$.



\end{document}